\documentclass[prb,floatfix,twocolumn,showpacs,superscriptaddress,amsmath,amssymb]{revtex4}

\usepackage{graphicx}
\usepackage{dcolumn}
\usepackage{bm}

\newcommand{\be}{\begin{equation}}
\newcommand{\ee}{\end{equation}}
\newcommand{\bea}{\begin{eqnarray}}
\newcommand{\eea}{\end{eqnarray}}
\newcommand{\nn}{\nonumber\\}

\newcommand{\pa}[1]{\left(#1\right)}
\def\eq#1{(\ref{#1})}
\def\v#1{{\bm#1}}

\begin{document}

\title{Influence of lattice distortions in classical spin systems}
\author{D.C.\ Cabra}
\affiliation{Laboratoire de Physique Th\'eorique, Universit\'e
Louis Pasteur, 3 rue de l'Universit\'e, F-67084 Strasbourg Cedex,
France} \affiliation{Departamento de F\'{\i}sica, Universidad
Nacional de la Plata, C.C.\ 67, (1900) La Plata, Argentina}
\affiliation{Facultad de Ingenier\'\i a, Universidad Nacional de
Lomas de Zamora, Cno. de Cintura y Juan XXIII, (1832) Lomas de
Zamora, Argentina.}
\author{M.\ Moliner}
\affiliation{Laboratoire de Physique Th\'eorique, Universit\'e
Louis Pasteur, 3 rue de l'Universit\'e, F-67084 Strasbourg Cedex,
France}
\author{F.\ Stauffer}
\affiliation{Laboratoire de Physique Th\'eorique, Universit\'e
Louis Pasteur, 3 rue de l'Universit\'e, F-67084 Strasbourg Cedex,
France}

\date{\today}

\begin{abstract}

We investigate a simple model of a frustrated classical spin chain
coupled to adiabatic phonons under an external magnetic field. A
thorough study of the magnetization properties is carried out both
numerically and analytically. We show that already a moderate
coupling with the lattice can stabilize a plateau at $1/3$ of the
saturation and discuss the deformation of the underlying lattice
in this phase. We also study the transition to
saturation where either a first or second order transition can
occur, depending on the couplings strength.
\end{abstract}


\maketitle

\section{Introduction and motivation}

The study of frustrated spin systems continues to be a subject of
intense research, in particular in low dimensions where the effect
of quantum fluctuations is more dramatic, leading to fairly rich
phase diagrams. On the one hand, one-dimensional frustrated quantum
spin systems are in general well under control, mainly thanks to the
availability of powerful techniques like bosonization \cite{oned}
and DMRG \cite{DMRG1,DMRG2,DMRG3}. On the other hand, these
techniques have unfortunately not been successfully generalized to
the two-dimensional case and there is then a strong need for the
development of useful techniques to analyze these systems
\cite{twod}.

A standard way to study quantum spin systems is to start from the
analysis of the classical (large $S$) limit and then try to include
the effects of quantum fluctuations in a systematic manner
\cite{Auerbach}. In certain cases, this procedure can lead to a
reasonable description of an otherwise intractable problem. Related
to this, the interplay between frustration and classical phonons has
been shown to lead to interesting features, even for the classical
spin system on the pyrochlore lattice \cite{Penc}, like the
stabilization of a magnetization plateau at $1/2$ of saturation.
Then, a natural question that arises is whether the classical limit
could be generally a good starting point to tackle the issue of the
interplay between frustration and lattice deformations and its
incidence on the appearance of magnetization plateaux. In the
present paper we analyze this point by focussing on a
one-dimensional $J_1-J_2$ model coupled to classical phonons, where
both the quantum and classical situations can be analyzed and
compared. This and related problems have been studied long time ago\cite{history1,history2,history3},
 but to our knowledge the magnetization
properties have not been analyzed so far.

The quantum version of this model has been studied in a recent
article \cite{magnetonos}, where it has been shown that the effects
of lattice distortions coupled to a given frustrated quantum spin
system can lead to new phases, in particular to plateaux and jumps
in the magnetization curve. Although plateaux phases are also
present in the pure spin system \cite{plateaux}, it has been shown
that lattice effects can lead to the enhancement of these phases
under certain circumstances. It is worthwhile mentioning that
inorganic compounds like
$\text{Cu}\text{Ge}\text{O}_3$\cite{coppergermanate} and
$\text{Li}\text{Vi}_2\text{O}_5$\cite{lithiumvanadate1,lithiumvanadate2}
are well described by the $J_1-J_2$ model, rendering its study both
theoretically and experimentally relevant. Values for the exchange
integrals,  such as $J_1\approx 160 K$ and the ratio $J_2/J_1\approx
0.36$, have also been proposed for copper germanate
\cite{coppergermanateexchange}.

We shall address the question of whether the
effects of these lattice deformations can already lead to
interesting magnetization properties at the classical level. The
main motivation for the present study is to analyze the origin of
such plateaux in the particular case of a classical zig-zag chain.
Although this case is particularly simple and the quantum model can
be treated using bosonization, understanding the role of lattice
deformations for classical spins could lead to a
way to study more involved situations, such as two dimensional
frustrated systems, where analytical techniques are not so
powerful than in one dimension as indicated earlier.

Let us consider the $J_1-J_2$ frustrated chain coupled to adiabatic
phonons
\begin{eqnarray}
\mathcal{H} &=&\frac{1}{2}K \sum_{i} \delta_{i}^2 + J_1\sum_{i}
(1-\tilde{A}_1\delta_{i})\, \v{S}_{i} \cdot \v{S}_{i+1}\nn
&&+ J_2
\sum_i\v{S}_i\cdot\v{S}_{i+2} -H \sum_i S^z_i.\label{ham}
\end{eqnarray}
%
In the previous hamiltonian, we chose to modulate only
the nearest neighbor (NN) interaction term,
and to consider there is no effect on the next to nearest neighbor
(NNN) coupling. This minimizes the number of parameters in the hamiltonian.
We have however checked that the inclusion of such a modulation on the
NNN couplings does not belie our main conclusions.

In the classical system phonons can be integrated out \cite{Kittel},
leading to an extra quartic interaction among the spins. The
effective Hamiltonian, written in units of $J_1$, reads

\begin{eqnarray}
\mathcal{H}_{\text{eff}} &=& \sum_i \pa{\v{S}_i\cdot\v{S}_{i+1} + \alpha
\v{S}_i\cdot\v{S}_{i+2}
- \frac{A_1^2}{2}
\pa{\v{S}_i\cdot\v{S}_{i+1}}^2}\nn&&-h\sum_i S^z_i,\label{hameff}
\end{eqnarray}
where the following reduced quantities $\alpha=
J_2/J_1$, $A_1=\frac{\tilde{A}_1}{K^{1/2}}$ and $h=H/J_1$ were defined. Even
though one ought to study the effect of the elastic constant $K$ and
$\tilde A_1$ separately, we will focus on the reduced coupling
$A_1$ whenever possible, reducing the number of parameters to a manageable size.

In Section \ref{sec:phasediagram} we study the magnetic phase
diagram using numerical and analytical techniques. We pinpoint a
region in the parameter space where a plateau appears at $M_z=1/3$
only. This should be contrasted with the quantum model, which shows
in addition a clear $M_z=0$ plateau in a wide region of the
parameter space, and another at $M_z=1/2$ in a narrower region.
Looking into the detailed structure of the ground state at these
plateaux, one can understand this discrepancy in the following way:
the structure at $M_z=1/3$ is of the ``Up-Up-Down''($UUD$) type,
indicating a classical plateau \cite{Affleck-Hida}, while in the
$M_z=0$ case the singlet structure can be identified with a quantum
one.

In Section \ref{sec:saturation}
we discuss the transition to saturation, which is
found to be either of first or second order depending on the
ratio between frustration and effective lattice coupling.

\section{$1/3$ magnetization plateau}\label{sec:phasediagram}

Let us analyze the magnetic phase diagram of the model
(\ref{hameff}). In the absence of an external magnetic field and
when $A_1 < \sqrt{4\alpha-1}$, the ground state is a spiral with a
pitch angle $\theta$  given by $\cos\theta = 1/(A_1^2-4\alpha)$. Its
energy is
\begin{equation}
E_{\text{spiral}} = \frac{1}{2}\cos\theta-\alpha.
\end{equation}
When  $A_1 > \sqrt{4\alpha-1}$ the ground state is N\'eel ordered.
The magnetization curves of this system show interesting features,
which vary depending on the relation between $\alpha$ and $A_1$,
as we discuss below.

\begin{figure}
\vspace{2mm}
\includegraphics[scale=0.85]{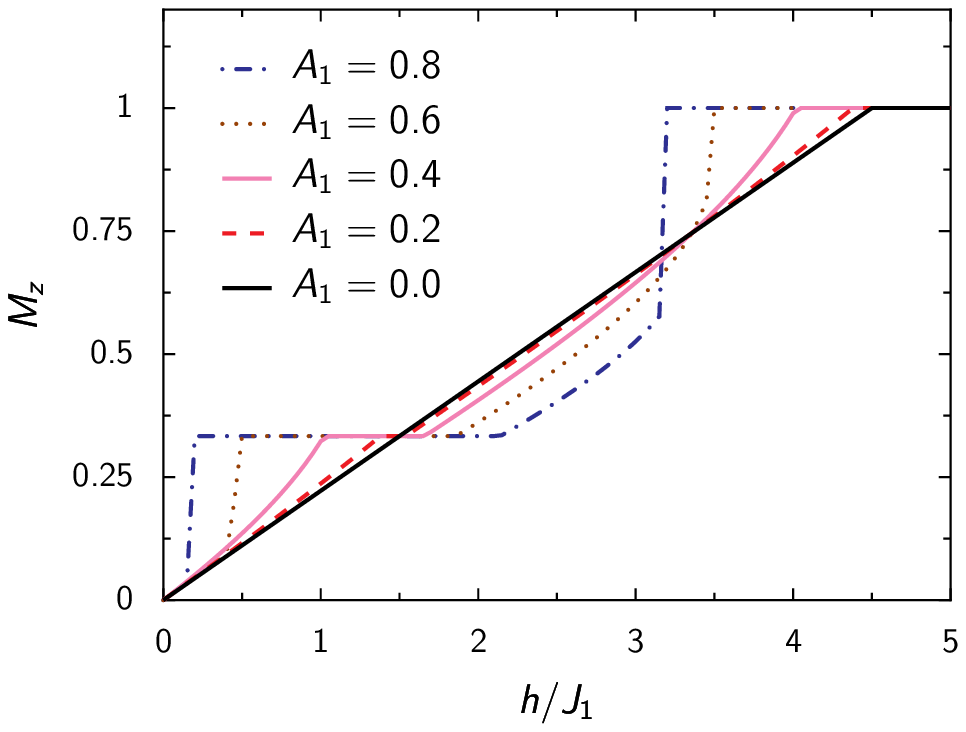}
\caption{\label{alphafixed} Magnetization curves $M(h)$ for
$N=30$ spins with $\alpha=0.5$ and ${A}_1=0.0,\cdots,0.8$ in
steps of $0.2$. Periodic boundary conditions are applied on
the chain.
The system is gradually cooled to $T\approx0$ over $3\times10^6$
Monte-Carlo sweeps (color version online). }
\end{figure}

In Fig.\ \ref{alphafixed} we represent $M(h)$ for a fixed value of
the frustration $\alpha=1/2$ and different values of the spin-phonon
coupling $A_1$. The data was obtained using classical Monte-Carlo
(MC) based on the usual Metropolis algorithm. Starting at
high-temperatures we perform several thousands of MC sweeps, and
then cool down the system to a fraction of the initial temperature.
This procedure is then repeated, slowly annealing the system to zero
temperature. We observe that a steady magnetization plateau at $1/3$
appears as soon as the coupling to the lattice is slightly turned
on, whose length increases with $A_1$. One can notice that the way
the system enters the plateau from the low-field side and eventually
saturates differs depending on the effective lattice coupling $A_1$.
For $A_1\gtrsim 0.6$ the two are first order transitions. Another
interesting characteristic seen in Fig.\ \ref{alphafixed}, is that
all curves represented (except one) cross at the same field
$h_{\times}\approx 3.35$ for which $M_{\times} = M(h_\times) \approx
0.745$. We shall discuss this point at the end of section
\ref{sec:saturation}. This brief overview suggests that the coupling
with the phonons stabilizes the state at $M_z=1/3$. Since the
plateaux are observed at zero temperature, we can fairly assume that
this effect is energy driven.

After this numerical preamble, it is time to derive some analytical
predictions on the characteristics of the magnetization plateau. For
this purpose, we need to find out which states describe the system
in the low and high-field regions around $M_z=1/3$. It is not
surprising that in the plateau phase the system adopts the $UUD$
state, but it will be a crucial point in our discussion as we shall
see later. In this state, the spins are aligned along the $z$-axis,
two up spins alternating with one down spin which is precisely the structure
seen at the $1/3$ plateau in the quantum model \cite{magnetonos}.

\begin{figure}
\vspace{2mm}
\includegraphics[scale=0.85]{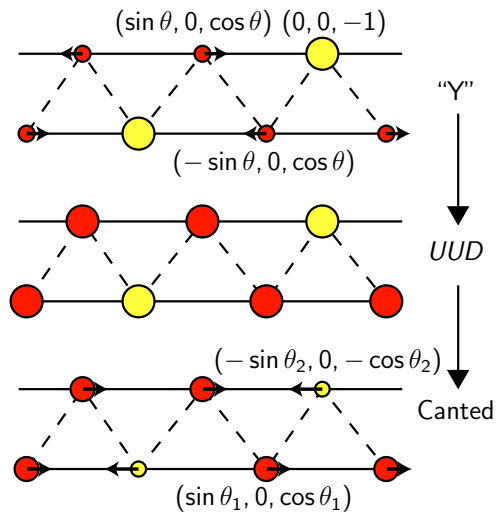}
\caption{\label{evol} Configurations observed in
the low-field, the 1/3 magnetization plateau and
high-field regions.
The chain is viewed in the $xy$-plane. The arrows
denote the projection of $\v{S}_i$ in this plane, whereas
the circles represent the $S_z$ component (red for $S_z>0$,
yellow otherwise and radius proportional to
$|S_z|$, color version online). The parametrization of the
states is given for each configurations. }
\end{figure}

The classical MC data indicates that the situation in the low-field
region is more complicated. On the one hand, the transition to the
$UUD$ state can occur at a very low field, where the system is not
far from its zero field ground state. Then, there is no small unit
cell structure providing a good description of the system, as the
spiral structure still prevails. On the other hand, when the
transition is smooth in the low-field region, a plausible assumption
is to consider that the system adopts a coplanar ``Y'' configuration
parameterized by a single angular degree of freedom $\theta$ (see
Fig.\ \ref{evol}). The unit cell energy for this state reads \bea
E_{Y}(\theta) &=& (1+\alpha)\pa{2\cos\theta(\cos\theta-1)-1}\nn
&&-\frac{A_1^2}{2}\pa{2\cos^2\theta+\pa{2\cos^2\theta-1}^2}\nn &&
-h(2\cos\theta-1). \eea
This expression can be minimized for any set of the parameters $h,\alpha$
and $A_1$. As the magnetic field increases, the solution will
eventually yield $\theta=0$ corresponding to the
$UUD$ state. This configuration is always a solution of $\partial_\theta
E_{Y}(\theta) = 0$, but it is only a minimum of the energy when
$h\geq h_Y = 1+\alpha-3{A_1}^2$. We should emphasize that this
discussion only makes sense whenever $h_Y$ is positive. For a
given value of the magnetic field, there can be other solutions
satisfying
\be h =
(1+\alpha)\pa{2\sqrt{1-X^2}-1}-{A_1}^2\pa{3-4X^2}\sqrt{1-X^2}\label{hy}.
\ee
where $X = \sin\theta$ (assuming $\cos\theta>0$).
The study of Eq.\ \eq{hy} boils down to finding the sign of a
polynomial expression. Introducing $\Delta =
2(1+\alpha)-11{A_1}^2$, we can show that when $\Delta\geq 0$ there
is exactly one more extremum of the energy for $h\leq h_Y$ and
that it is always a minimum. This solution becomes precisely the
$UUD$ state at $h=h_Y$. Under these assumptions, we can conclude
that the critical field for which we recover $M_z = 1/3$ from the
low field regime is

\be h_{c1} = 1+\alpha-3{A_1}^2,~~~\Delta \geq 0. \label{hc1} \ee

This can be compared to our MC results. For instance, the data for
$\alpha=0.5$ and $A_1=0.4$ (solid pink curve in Fig.
\ref{alphafixed} ) allows us to obtain a precise estimate for
$h_{c1}$ at $T\approx0$. We get $h_{c1} = 1.02 \pm 0.01$. For this
set of parameters, $\Delta$ is positive so that we are ruled by the
previous assumptions. The analytical expression \eq{hc1} yields
$h_{c1} = 1.02$, which is in excellent agreement with the
simulations. 

For $\Delta<0$ there can be up to two extra solutions when $h\geq
h_Y$. As there is always one solution that never turns out to become
$UUD$ for a certain value of the magnetic field, we ought to perform
a detailed comparison of the two solutions' energies in order to
conclude. We shall not step further into this discussion, which can
nevertheless be conducted numerically using the previous analytical
expressions. For instance we performed it when $\alpha=1/2,
A_1=0.6$, leading to $h_{c1}\approx 0.46$. This is in good agreement
with the MC data which gives $h_{c1}\approx 0.47\pm 0.01$  (dotted curve in Fig.\ref{alphafixed} ). It can be
understood from the previous discussion that $h_Y$ is always a lower
boundary of the critical field: \be h_{c1}\geq
1+\alpha-3{A_1}^2,~~\Delta<0. \ee

If we increase $A_1$ while keeping $\alpha$ fixed, $h_Y$ eventually
becomes negative (as it is the case for $A_1 = 0.8$, dash-dotted
blue curve in Fig. \ref{alphafixed}) and we can generally not
conclude using this small unit cell configuration. The reader should
keep in mind that the regime where $A_1$ becomes large is not well
described by our initial hamiltonian (1) since in that case one
should include the effects of the lattice also in the NNN
interactions.

\begin{figure}
\vspace{2mm}
\includegraphics[scale=0.85]{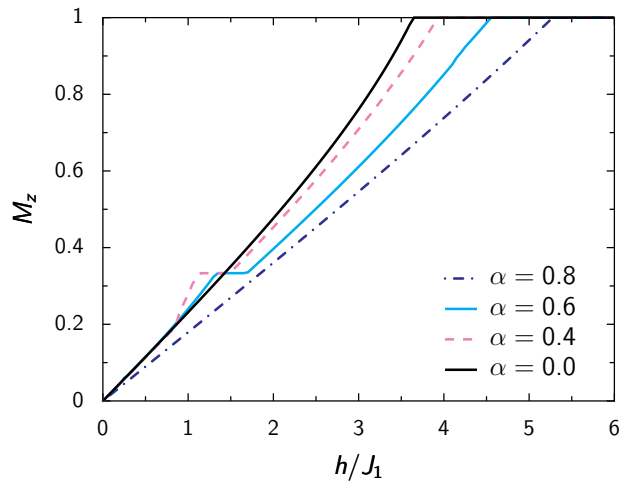}
\caption{\label{a1fixed} Magnetization curves $M(h)$ for $N=30$ spins
with $A_1=0.3$ and ${\alpha}=0.0, 0.4, 0.6, 0.8$. Periodic boundary
conditions are applied on the chain.
The system is gradually cooled to $T\approx0$ over $3\times10^6$
Monte-Carlo sweeps (color version online). }
\end{figure}

We shall now focus on the state observed in the high-field region to
find the corresponding upper critical field $h_{c2}$ above which the
plateau disappears. By comparing $h_{c1}$ to $h_{c2}$, we should be
able to conclude on the existence of the $1/3$ magnetization
plateau. In the upper critical region, the situation is far more
under control. The system can be seen to be well described by a
3-spins coplanar ``canted'' configuration with two degrees of
freedom (see Fig.\ \ref{evol}). The energy of such a configuration
is given by

\begin{eqnarray}
    E_{\text{canted}}(\theta_1,\theta_2) & =  & (1+\alpha)\pa{1 - 2U(\theta_1,\theta_2)}\nn
    &&-\frac{A_1^2}{2}\pa{1+2U(\theta_1,\theta_2)^2}\nn
    &&-h(2\cos\theta_1-\cos\theta_2)\label{ecanted},
\end{eqnarray}
where $U(\theta_1,\theta_2) = \sin\theta_1\sin\theta_2+\cos\theta_1\cos\theta_2$.

The configuration $UUD$, which corresponds to
$\theta_1=\theta_2=0$, is always a critical point of the function
$E_{\text{canted}}(\theta_1,\theta_2)$. A closer look at the
second order derivatives with respect to $\theta_1$ and $\theta_2$
shows that it is a local minimum only for $ 0 \leq h \leq
1+\alpha+{A_1}^2$. The other critical points satisfy the following
set of equations
\bea
 Y & = & 2X\label{eqsin}, \\
 h & = & \pa{1+\alpha+A_1^2\pa{2X^2+\sqrt{1-X^2} \sqrt{1-4X^2}\sigma_1\sigma_2 }}\nn
 &&\hspace{0.5cm}\times\pa{2\sqrt{1-X^2} - \sqrt{1-4X^2}\sigma_1\sigma_2}\label{eqfield},
\eea
where
\begin{eqnarray}
    \sin\theta_1 & =&  X, \cos\theta_1  =  \sigma_1\sqrt{\pa{1-X^2}},\\
    \sin\theta_2 & = &  Y, \cos\theta_2 = \sigma_2\sqrt{\pa{1-Y^2}}.
\end{eqnarray}
The quantities $\sigma_1,\sigma_2=\pm 1$ account for all the possible signs of
both cosines. We see from Eq. \eq{eqsin} that there is a strong
constraint on $(\theta_1,\theta_2)$ verified regardless of
the values of the couplings. At $h=1+\alpha+{A_1}^2$, Eq.
\eq{eqfield} admits only one solution which turns out to be $UUD$.
For larger value of $h$, $UUD$ can no longer be a critical point,
which implies
\be
h_{c2} = 1+\alpha+{A_1}^2,
\ee
corresponding to the exit of the plateau in the high-field region.
From this discussion it can be concluded that whenever our
assumptions are correct, there is a plateau at $M_z=1/3$ of length
$\Delta h_{1/3} = 4A_1^2$ starting at $h_{c1}$. This result has been
checked to be consistent with the MC computations and the analytical
value of $h_{c2}$ matches the value estimated from all the curves in
Fig. \ref{alphafixed}.

\begin{figure}
\vspace{2mm}
\includegraphics[scale=0.85]{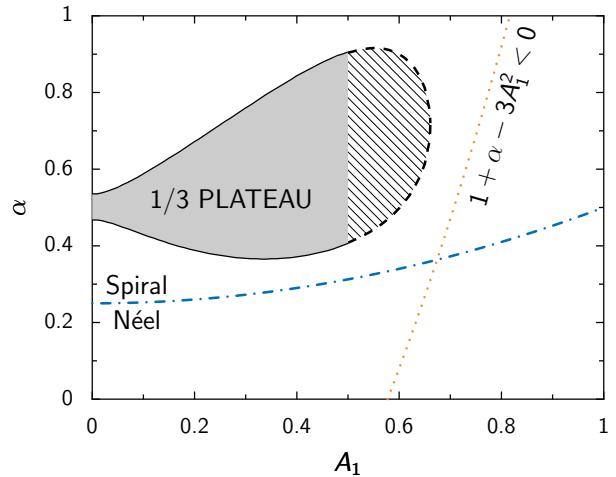}
\caption{\label{phased} Qualitative  $(A_1,\alpha)$ phase diagram (color version online).
The filled area between the two full curves corresponds to the
region of the parameters space where the $M_z=1/3$ plateau is observed.
The hatched region between the two dashed curves is the
region of the parameter space for which our
approach is no longer observed to be fully valid. Also represented
the limit between N\'eel and Spiral ground states at $h=0$ (blue dash-dotted
line) and the region where $1+\alpha-3A_1^2<0$ (dotted orange
line).}
\end{figure}

There is one more question we need to address: for which set of
parameters $(A_1,\alpha)$ can we observe this plateau? Under the
previous assumptions regarding the states observed in the low and
high-field regions, we can conclude it exists for any $A_1>0$. Yet
the system can not be described in such a manner for all values of
$\alpha$ and $A_1$. Working at a fixed lattice coupling $A_1=0.3$,
we were able to obtain some magnetization curves varying the
frustration $\alpha$. Some of those curves are plotted in Fig.
\ref{a1fixed}, which clearly shows that there is only a narrow
region in $\alpha$ where the plateau is observed. A precise answer
to the previous question is rather challenging, and we shall first
try to discuss this point in a more qualitative manner before
adopting a more precise strategy. At $M_z=1/3$, we can of course
expect to see a lot of different configurations, depending on the
values of the couplings. However, the MC simulations suggest that
the plateau always corresponds to the $UUD$ configuration. This
state is perfectly collinear, minimizing the quartic contribution to
the effective hamiltonian \eq{hameff}. For instance, it can be seen
numerically that for $\alpha=1/2$ with no coupling to the lattice,
the system reaches $1/3$ magnetization in the $UUD$ configuration.
Even a small positive value of $A_1$ will then stabilize the $UUD$
state enough for it to be stable when the field is slightly
increased. On the opposite, if one antiferromagnetic coupling
dominates the other, the system will be in a different state at
$M_z=1/3$. In the extreme case where $\alpha\approx 0$ for instance,
the system will favor N\'eel order in the $xy$ plane, each spin
having the same $z$-axis projection $S_z=1/3$. This layout already
trades off some collinearity in favor of magnetic field alignment.
There is no surprise that this trade-off will be further enhanced as
the magnetic field is increased, so that no plateau should be
observed.

\begin{figure}
\vspace{2mm}
\includegraphics[scale=0.85]{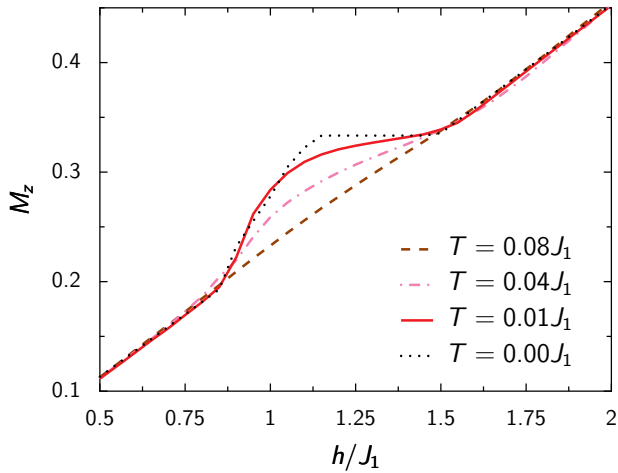}
\caption{\label{alphaa1fixed} Magnetization curves $M(h)$ for
$N=30$ spins with $A_1=0.3$ and ${\alpha}=0.4$ around  $M_z = 1/3$
for different temperatures $T=0,0.04,0.08$ in units of $J_1$.
Averages are computed
on $2^{22}$ sweeps through the lattice after an initial $2^{18}$ sweeps of
thermalization. Increasing the temperature quickly destroys the
plateau observed at $M_z=1/3$. }
\end{figure}

A more accurate way to tackle this issue is to start from the
$h=0$ spiral ground state and ponder over the state adopted by the
system when the magnetic field increases. We have already
performed part of this task earlier, suggesting that the system
slowly moves to a ``Y'' configuration, whose out of plane
components make it a ``precursor'' of the $UUD$ configuration. Another
plausible solution is that the spins, while keeping their spiral
structure in the $xy$-plane, all acquire the same $S_z$ projection. In
this case, the $n^{\text{th}}$ spin reads
\begin{equation}
    \v{S}_{n} = \pa{\sqrt{1-z^2}\cos(n\theta),\sqrt{1-z^2}\sin(n\theta),z},
\end{equation}
with $\cos\theta = 1/(A_1^2-4\alpha)$. The energy per site,
\begin{eqnarray}
    E_{\text{sz}}(z) &=& z^2+\left(1-z^2\right)
   \cos\theta+\alpha
   \left(z^2+\left(1-z^2\right) \cos
  2 \theta \right)\nn
&&-\frac{A_1^2}{2}
       \left(z^2+\left(1-z^2\right) \cos
       \theta
       \right)^2-h z,
\end{eqnarray}
can be minimized with respect to $z$ to find the lowest energy
configuration at a given magnetic field. Our idea is to perform this
minimization at $h=h_{c1}$, and to see if the corresponding
configuration is of lower energy than UUD at the same field. If so,
the system will not enter the plateau at $h_{c1}$, and of course as
mentioned the previous paragraph no plateau should be observed. For
a fixed value of $A_1$, we can determine the range in $\alpha$
leading to $UUD$ at $h_{c1}$. The roots of the polynomial equation
are evaluated numerically, from which we sketch the phase diagram
represented in Fig. \ref{phased}. This approach only makes sense
when we have a precise value for $h_{c1}$, which we saw is the case
if $A_1$ is not too large ($A_1\lesssim 0.5$ from the MC data). We
notice that the diagram is in agreement with the situation depicted
in Fig. \ref{a1fixed}, as well as the one in Fig. \ref{alphafixed}
when $A_1$ is not too large. The most remarkable feature is that for
an arbitrary small yet strictly positive $A_1$, one can find a value
of $\alpha$ for which the plateau phase is observed.

The effect of temperature on the magnetization plateau is
potentially important as an ``order by disorder''
effect\cite{Villain} could further stabilize the plateau. We
investigated this point by performing our MC simulations at different
temperatures, without annealing the system. 
A sample is given in Fig. \ref{alphaa1fixed}, and in
general we observed no remarkable features. The increasing thermal
fluctuations quickly destroy the plateau. We should also mention
that we observed no strong finite size effects in the numerical
simulations, which is why we were always able to work on systems
with less than a hundred spins.

\begin{figure}
\vspace{2mm}
\includegraphics[scale=0.85]{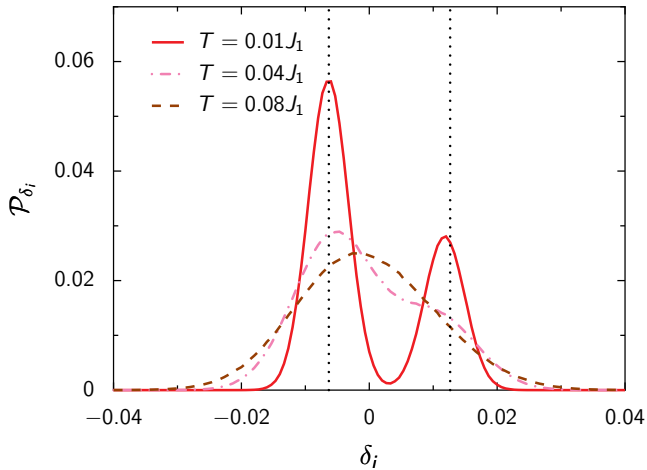}
\caption{\label{latticedistribution} Normalized lattice displacement
histograms   for $N=30$ spins with $\alpha = 0.5$, $A_1=0.4$ ($K = 10^3 J_1$)
and $h=1.5$ at different temperatures. $300$ points were used in the
interval $[-0.1,0.1]$. The two dotted vertical lines correspond to
the $T=0$ limit calculated in the text. The data was obtained using
a direct classical Monte-Carlo for the hamiltonian \eq{ham}.   }
\end{figure}

We conclude this section by focusing on the lattice deformations.
Until now, we studied the effective spin-only hamiltonian
\eq{hameff} which embodies the straightforward analytical approach
to the problem. It is however important to get more insight on the
structure of the lattice deformation inside the plateau phase. For
that matter, we modified our MC algorithm to take into account the
lattice degrees of freedom as well. Starting from the hamiltonian
\eq{ham}, we used the Metropolis algorithm for both the spin
positions and orientations, applying periodic boundary conditions on
the chain. We studied the normalized histograms of the displacements
$\delta_i$  at finite temperature. We fixed $\alpha = 0.4$ and $A_1
= 0.3$, the same values used in Fig. \ref{alphaa1fixed} to allow a
direct comparison between the two figures, and selected the magnetic
field so that the system is at $M_z\approx 1/3$. Besides the value
of $A_1 = \tilde{A}_1/\sqrt{K}$, we need to give $K$, the spring
constant in \eq{ham}, a sensible value. We took $K=10^3J_1$, large
enough to make sure the displacements remain small. This corresponds
to $\tilde{A}_1\approx 9.5$. We mention that both $K$ and $A_1$ are
of the same order of magnitude as the one for a more complex two
dimensional material such as
$\text{Sr}\text{Cu}_2{(\text{BO}_3)}_2$\cite{SS} and that they can
be considered at least as ``realistic'' for copper germanate or
lithium vanadate\cite{Becca}. The results are given in Fig.
\ref{latticedistribution}.

We see that the lattice deformations are not uniform and that their
histogram presents two peaks at $T=0.01J_1$. They are centered
around a negative and positive value of the displacement $\delta_i$.
This suggests that the underlying deformation consists of ``UDU''
trimers (Up-Down-Up) on the chain. Let us introduce $\delta_+$ the
displacement between two consecutive trimers and $\delta_-$ the
displacement between the down spin and its two nearest neighbors
inside the trimer. The energy of this unit cell is given by:
\be
    E = 3K\delta_{-}^2 + 4J_1\tilde{A}_1\delta_{-} -J_1 +J_2 - H,
\ee
where the periodic boundary conditions imply
$\delta_{+}=-2\delta_{-}$. Minimizing the energy, the deformation
should become

\be
    \delta_{+}=-2\delta_{-}= \frac{4\tilde{A}_1J_1}{3K}
\ee
at $T=0K$. Going back to Fig. \ref{latticedistribution}, at
$T=0.01J_1$ the distribution clearly exhibits two peaks and we can see
that they are almost centered around $\delta_{+}$ and $\delta_{-}$
respectively. The ratio between the height of the two peaks is about
$2$, a consequence of the fact there are twice as many up spins than
down spins in the $UUD$ state. Those results seem to validate the trimer scenario at
low temperature. When the temperature increases to $T=0.04J_1$, the
peaks start to overlap, betraying the gradual destruction of the
plateau already seen in Fig. \ref{latticedistribution}. Finally at
$T=0.08J_1$, we end up with a single peaked, almost gaussian,
distribution: the plateau eventually disappeared. We end up by
stating that expectation value of the displacement is always zero as
the periodic boundary conditions applied ensure the length of the
chain remains fixed throughout the simulation.

\section{Transition to saturation}\label{sec:saturation}

The study of the upper critical magnetic field yields another
interesting result: we can get a precise picture of how the system
eventually reaches saturation. This result can be foreseen using
classical MC, which shows that the canted state describes the
system quite well even for $h>h_{c2}$. A close look at Fig.\
\ref{alphafixed} shows that two different behaviors of the
magnetization between $h_{c2}$ and the saturation value are
observed. For different values of spin-phonon coupling, the system
can undergo a first or second order transition to reach
saturation. We are going to demonstrate that this result can be derived from energetic
considerations on the canted state. From Eq. \eq{ecanted}, we see
that the saturated state, reached for $\theta_1=0$ and
$\theta_2=\pi$, minimizes the energy for a magnetic field greater
than $h_{c3}=3(1+\alpha-{A_1}^2)$. This imposes a lower boundary on the
saturation field $h_{U}$. We assume that the couplings $A_1$ and
$\alpha$ are such that $h_{c2} < h_{c3}$, a situation where the
previous discussion on the existence of the $1/3$ magnetization
plateau still holds.  To be consistent with the state of system for
$h>h_{c2}$, we set $\sigma_1 = 1$ and let $\sigma_2=-\sigma$ take
the values $\pm 1$ so as to be able to move from $UUD$ to
saturation continuously. At a given magnetic field, one can obtain
the corresponding critical configurations by finding the roots of
Eq. \eq{eqfield}. This task reduces to the study of the two functions
$h_{\sigma}$:
\bea
 h_\sigma(X) & = & \pa{1+\alpha+A_1^2\pa{2X^2-\sqrt{1-X^2} \sqrt{1-4X^2}\sigma }}\nn
 &&\hspace{0.5cm}\times\pa{2\sqrt{1-X^2} + \sqrt{1-4X^2}\sigma}\label{hsigmadef}.
\eea

\begin{figure}
\vspace{2mm}
\includegraphics[scale=0.75]{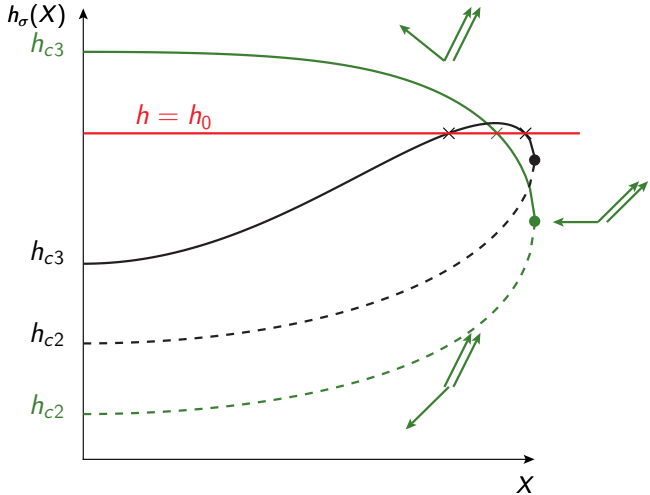}
\caption{\label{h_sigma} ``Low magnetization'' (dashed line) and ``high
magnetization'' (plain line) curves as a functions of
$X=\sin\theta_1$ for $\alpha=0.5$, $A_1=0.5$ (green) and
$\alpha=0.5$, $A_1=0.8$ (black). The typical shape of the minima's
unit cell along the green curves are depicted in the different magnetic field regions.
The intersections (crosses) with the line $h=h_0$ (red) gives
the competing critical point at this magnetic field. The colored dots
correspond to the minimal configuration with one spin in the $xy$-plane.}
\end{figure}

Their roots can be determined graphically for a fixed field $h_0$ as
they are the values of $X$ for which the line $h=h_0$ intersects
$h_\sigma(X)$. The ``low magnetization'' function $h_{-}$ will give
us solutions with one spin still pointing down, whereas the ``high
magnetization'' function $h_{+}$ will give us states where all the
spins have a positive $S_z$ component. Fig.\ \ref{h_sigma} is a plot
of both functions for two sets of values $\alpha$, $A_1$. In both
cases, the curves for $h_{+}$ and $h_{-}$ join at
$h_{c4}=\sqrt{3}(1+\alpha+A_1/2)$ (colored dots in  Fig.\
\ref{h_sigma}). For this value of the magnetic field, the root of
$h_{\sigma}$ corresponds to a configuration in which one of the
three spins lies precisely in the $xy$-plane. Two possible behaviors
are observed. For instance when $\alpha=A_1=0.5$ (green curves in
Fig.\ \ref{h_sigma}), we see that for a fixed magnetic field
$h\in[h_{c2},h{c3}]$ there is only one critical point of the energy,
which can be shown to be a minimum. We are able to follow easily the
state of system as the magnetic field increases. The two up spins
first slightly tilt to let the down spin reach the $xy$-plane and
then they all progressively align along the $z$-axis while still
satisfying Eq. \eq{eqsin}. The three-spins unit cell configuration
smoothly goes from
 $UUD$ to saturation.

\begin{figure}
\vspace{2mm}
\includegraphics[scale=0.85]{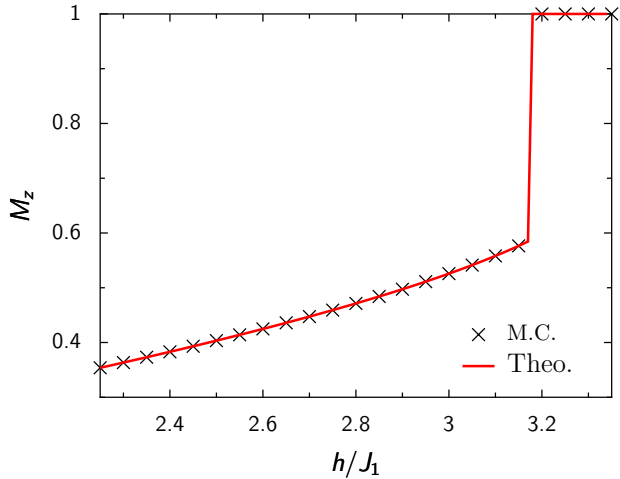}
\caption{\label{theovsmc} Upper region of the magnetization curve
for $\alpha = 0.5$ and $A_1=0.8$ at $T=0$. The full line
represents the results obtained using the analytical minimization
of the ``canted'' state  energy, crosses the data obtained by
Monte-Carlo. }
\end{figure}

For $A_1=0.8$ (black curves in Fig.\ \ref{h_sigma}), the ``high
magnetization'' function $h_{+}$ (plain line) presents a maximum. In
this case, three states are potentially competing for $h$ between
$h_{c3}$ and its maximum value: the saturated state and the two
roots of $h_{+}$. We ought to compare their energies to conclude,
but it is not surprising that the outcome can be a first order
transition to saturation. We numerically solved the analytical
equations involved to get the magnetization curve from the exit of
the plateau to saturation for $A_1=0.8$. The saturation field we obtain is
$h_{U} \approx 3.17063$, for which the system jumps from $M_z\approx
0.58434$ to saturation. The comparison between this minimization and
the MC data is given in Fig. \ref{theovsmc} and shows the excellent
agreement achieved.

A more in-depth study of the  $h_\sigma$ functions' extrema allows
to work out the range in $(\alpha,A_1)$ for which
the transition to saturation is of first or of second order.
The former, which are related to the existence of a non trivial
maximum in $h_{+}$, occur only if $1 \leq
\pa{1+\alpha}/A_1^2\leq \frac{11}{2}$ in agreement with our numerical
observations.

Finally, it can be pointed out from Eq. \eq{hsigmadef} that for
$X=1/\sqrt{5}$ the function $h_{+}$ does no longer depend on the
coupling $A_1$. This state (if reached) will be the minimum
of the canted configuration energy for a magnetic field
\be
h_{\times} =\frac{5}{\sqrt{5}}(1+\alpha).
\ee

At this field the magnetization is $M_z= M_{\times} =5/(3\sqrt{5})$.
This explains why for our selection of parameters, all the curves
except one in Fig. \ref{alphafixed} cross at a field whose estimate,
given in Section \ref{sec:phasediagram}, coincides with
$h_{\times}$. Regardless the value of $A_1$, if the system is not
saturated at $h_{\times}$ then its magnetization will always be
$M_{\times}$.

\section{Summary and Conclusion}

The effect of lattice deformations at the classical level in a
frustrated spin system has been illustrated working on a simple
$J_1-J_2$ spin chain coupled to adiabatic phonons. We provide an
overall picture of the magnetization properties for a large set of
the parameters $\alpha,A_1$ introduced in our model. We have found
that a plateau at $M_z=1/3$ is present in certain region of the
parameters space, while no other plateaux are observed. Frustration
is a necessary ingredient, as the plateaux can only arise when the
zero field ground state is a spiral. The other ingredient, the
coupling to lattice deformations, is such that for an arbitrary
small yet strictly positive $A_1$, one can find a value of $\alpha$
for which the plateau phase is beheld. Further increasing $A_1$ will
broaden the region in the parameter space for which the plateau
occurs,  until the effective coupling is no longer mild enough for
our analytical approach to be valid, even if a numerical approach is
still achievable. It should be emphasized that the  stabilization
mechanism is purely energy driven and  triggered by the quartic
interaction induced by the lattice coupling in the effective
hamiltonian \eq{hameff}. The underlying lattice deformation shows
the chain is made of ``UDU'' trimers inside the plateau phase.


The absence of plateaux at $M_z=0$ and $M_z=1/2$ in the classical
model as compared to the quantum case can be understood by
analyzing the ground state structure of the plateaux in the
quantum case. It is only for $M_z=1/3$ that one observes a classical
type of spin configuration, of the $UUD$ type, while in the other
cases a quantum state is apparent.

\acknowledgments We thank A. Dobry, A. Honecker, P. Holdsworth and
T. Vekua for helpful discussions. This work was partially
supported by ECOS-Sud Argentina-France collaboration (Grant
A04E03) and PICS CNRS-CONICET (Grant 18294).

\end{document}